\documentclass[10pt,twocolumn,letterpaper]{article}

\usepackage{cvpr}
\usepackage{times}
\usepackage{epsfig}
\usepackage{graphicx}
\usepackage{amsmath}
\usepackage{amssymb}

\usepackage[breaklinks=true,bookmarks=false]{hyperref}

\cvprfinalcopy 

\def\cvprPaperID{****} 

\newcommand{\martim}[1]{#1}

\begin{document}

\title{Age and gender bias in pedestrian detection algorithms}

\author{Martim Brand\~ao\\
Oxford Robotics Institute\\
University of Oxford\\
{\tt\small martim@robots.ox.ac.uk}
}

\maketitle

\begin{abstract}
Pedestrian detection algorithms are important components of mobile robots, such as autonomous vehicles, which directly relate to human safety. Performance disparities in these algorithms could translate into disparate impact in the form of biased accident outcomes.
To evaluate the need for such concerns,
we characterize the age and gender bias in the performance of state-of-the-art pedestrian detection algorithms. 
Our analysis is based on the INRIA Person Dataset extended with child, adult, male and female labels. We show that all of the 24 top-performing methods of the Caltech Pedestrian Detection Benchmark have higher miss rates on children. The difference is significant and we analyse how it varies with the classifier, features and training data used by the methods. Algorithms were also gender-biased on average but the performance differences were not significant.
We discuss the source of the bias, the ethical implications, possible technical solutions and barriers.
\end{abstract}

\section{Introduction}

Recent research has brought to light some problematic biases in computer systems \cite{Friedman1996}, artificial intelligence (AI) \cite{Bolukbasi2016,Chouldechova2017} and robotics \cite{Howard2018}. These studies, together with ethical and social studies of disparate impact \cite{Barocas2016}, the nature of algorithm discrimination \cite{Boyd2014}, and concrete algorithm audits \cite{Bolukbasi2016,Chouldechova2017,Buolamwini2018,Wilson2019}, have shown the existence of a fairness and justice dimension of algorithms.
In computer vision, recent papers have audited gender classification \cite{Buolamwini2018}, object detection \cite{Wilson2019} and facial analysis algorithms
\cite{Taati2019}. Consistently, these audits have found troubling biases in the performance of algorithms.

In this paper we focus on pedestrian detection algorithms. The motivation for this audit is its relevance for the \emph{physical} safety of the people involved. Pedestrian detection algorithms are used by mobile robots such as autonomous vehicles (AVs), and performance differences in such algorithms could lead to disparate impact in the form of biased crash outcomes.
Recent examples such as the Uber accident due to the low-confidence detection of a pedestrian \cite{Uber2018} further ground our motivation, since depending on training set or algorithm biases these low-confidences (or missed detections) could be more likely to happen on specific kinds of pedestrians. 
Here we particularly focus on age and gender bias evaluation---and conclude that there is a clear worse performance of state-of-the-art algorithms on children. 
We complement the analysis with some discussion on the source of the bias, ethical implications, possible solutions, and barriers to ``solving'' the issue.

\section{Method}

\begin{figure*}
	\begin{center}
		\includegraphics[width=0.49\linewidth]{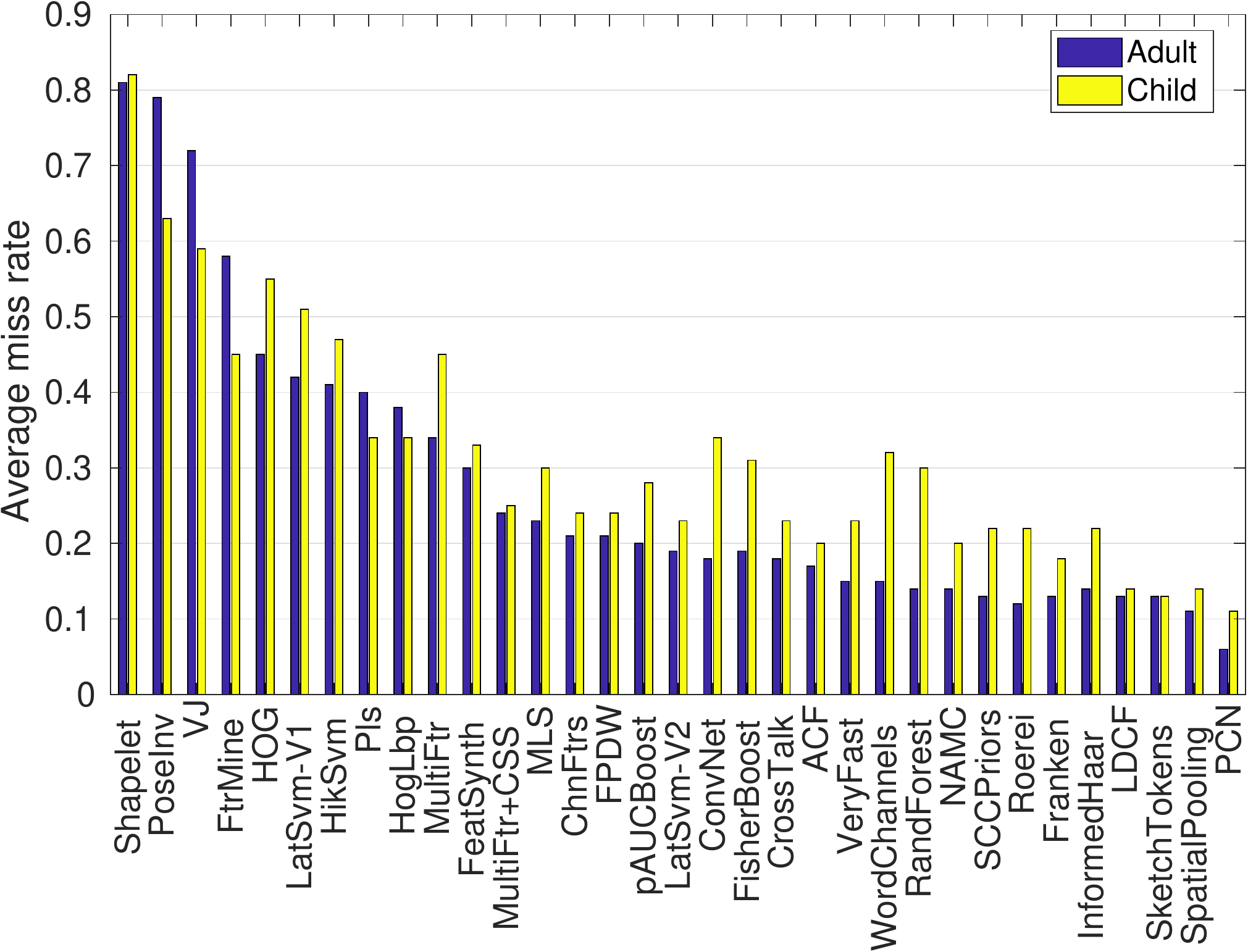}
		\includegraphics[width=0.49\linewidth]{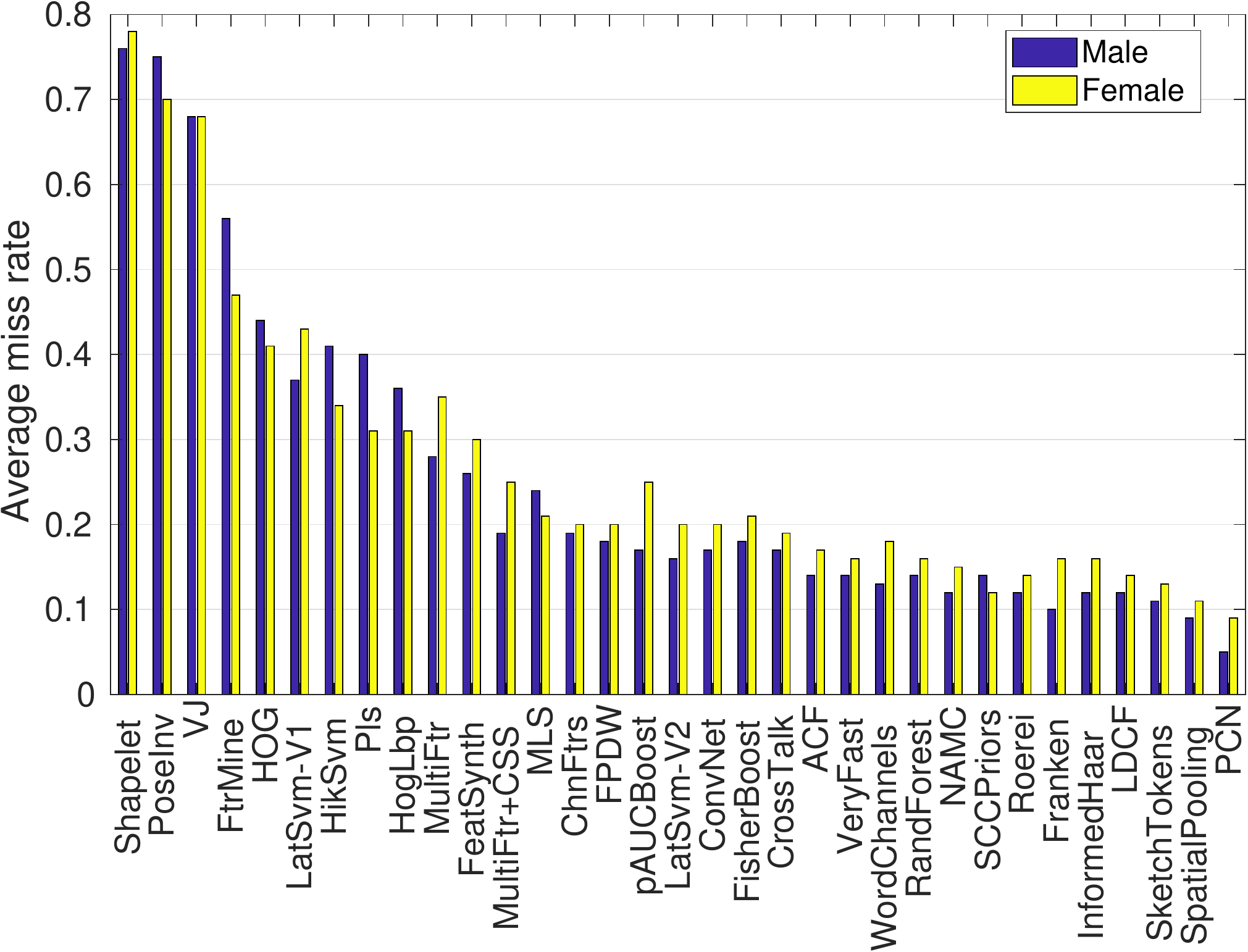}
	\end{center}
	\caption{Average miss rates of all methods available within the Caltech Pedestrian Detection Benchmark, evaluated on child, adult, female and male-labelled bounding-boxes only.}
	\label{fig:ACMF}
\end{figure*}

\subsection{Dataset}

Our dataset is an extended version of the INRIA Person Dataset \cite{Dalal2005}. The INRIA dataset is commonly used for the evaluation of pedestrian detection algorithms in the literature, for example within the comprehensive Caltech Pedestrian Detection Benchmark \cite{Dollar2012}. The dataset consists of pictures of pedestrians in streets and other urban scenarios, annotated with bounding boxes of the pedestrians.
We extended the test set of the INRIA dataset with gender and age labels for each bounding box. In particular, we extended each bounding box with \martim{an annotator-ascribed} gender label which can be either ``male'' or ``female'', and an age label which can be either ``child'' or ``adult'' \martim{(see last section for a discussion)}. We use the biological definition of ``child'', as a human between birth and puberty.
Our assumption when deciding to add this label was that children could be more difficult to detect by algorithms because of their size (smaller bounding boxes are related to lower detection rates \cite{Dollar2012}) and because they are typically more likely to use varied postures (e.g. bending down to grab something of interest, running, sitting). We did not expect to see differences in gender due to the nature of the task (i.e. to detect a full-body pedestrian), but decided we should do it as a way to contrast gender-bias issues in face \cite{Buolamwini2018,Taati2019} and pedestrian detection algorithms.
The labelling process was manual and done by a single annotator---the author. All pictures were seen in order, and for each picture the annotator ticked male/female/adult/child checkboxes placed over each of the picture's ground-truth bounding boxes. The final labels will be made available at {\footnotesize\url{http://www.martimbrandao.com/}}.

\subsection{Measuring bias}

The Caltech Pedestrian Detection Benchmark \cite{Dollar2012,CaltechPedestrianDetectionBenchmark} provides detection results for various competitive and state-of-the-art algorithms on the test-set of the INRIA dataset. We used the open results and evaluation code from this benchmark to evaluate method performance on subsets of the dataset. In particular, we evaluated algorithm performance in \martim{the ``reasonable'' subset of the test-set\footnote{Corresponding to a near view and no or partial occlusion. This is used to disentangle between age/gender and small-bounding-box effects, but it is also the main subset used in \cite{Dollar2012} for evaluation.}}, as well as male-, female-, child- and adult-only portions of that subset.
\martim{There are 589 bounding box detections in this set in total, of which 37\% were labelled ``female'' and 7\% ``child''. This already shows a bias at the dataset level.}
We use \emph{average miss rate} as the algorithm performance metric\martim{, where a miss is an area overlap below 50\%,} as in \cite{Dollar2012}. In the context of this paper, this metric is also an intuitive indicator of safety risk---for example as a pedestrian being ``missed'' can directly translate into a collision with a mobile robot such as an autonomous vehicle \cite{Uber2018}.

We evaluate bias qualitatively by comparing male/female and child/adult miss rates, as well as quantitatively by computing the average performance differences, ratios and Wilcoxon rank-sum test p-values. Similarly to the analysis of \cite{Taati2019}, we use the rank-sum test as a measure of whether the distribution of performance is the same for female/male and child/adult consistently over multiple algorithms or not.

To evaluate the algorithms on subsets of bounding boxes we used the ``ignore regions'' methodology of \cite{Dollar2012} and as implemented in \cite{CaltechPedestrianDetectionBenchmark}. In this methodology, subsets of boxes (i.e. male, female, child or adult) are ``ignored'' by not requiring detections on those boxes and not considering bad detections as mistakes either%
\footnote{In \cite{Dollar2012} the methodology is used to study the influence of bounding box size and occlusion on algorithm performance.}. 
We evaluate the performance on children by ignoring adults, on adults by ignoring children, etc. For simplicity, our labelling system does not allow simultaneous male-and-female or child-and-adult labels.

\section{Results}

Figure~\ref{fig:ACMF} shows the average miss rates for the child, adult, female and male subsets of the dataset. 
The miss rates were higher for children on 27 methods out of 33 (82\%). This percentage goes to 100\% when considering only the 24 best performing methods (miss rates below 30\%). On average, the difference between adult and children's miss rates was 0.04 percentage points, or 0.07 for the 24 best performing methods. These differences correspond to 1.3 times higher miss rates for children than adults averaged over all methods, or 1.4 when averaged over the top 24 methods.
Regarding gender, miss rates were higher for female pedestrians on 24 methods out of 33 (72\%). On average the difference was 0.01 percentage points (1.1 ratio).

\begin{table*}
	\begin{center}
		\begin{tabular}{|l|c|c|c|c|c|c|c|}
\hline
Method classifier 	 & \#methods 	 & Avg diff 	 & Avg diff 	 & Avg ratio 	 & Avg ratio 	 &   Rank-sum 	 &  Rank-sum \\
&           	 & Age 	 & Gender 	 & Age 	 & Gender 	 &   Age 	 &  Gender \\
\hline\hline
AdaBoost 	 & 19 	 & 0.03 	 & 0.01 & 1.29 	 & 1.13 	 &     0.063\hphantom{*} 	 &     0.213\hphantom{*} \\
deep net 	 & 2 	 & 0.11 	 & 0.03 & 1.86 	 & 1.49 	 &     0.667\hphantom{*} 	 &     0.667\hphantom{*} \\
latent SVM 	 & 2 	 & 0.07 	 & 0.05 & 1.21 	 & 1.21 	 &     0.667\hphantom{*} 	 &     0.667\hphantom{*} \\
linear SVM 	 & 4 	 & 0.03 	 & 0.00 & 1.06 	 & 1.07 	 &     0.886\hphantom{*} 	 &     1.000\hphantom{*} \\
pAUCBoost 	 & 2 	 & 0.06 	 & 0.05 & 1.34 	 & 1.35 	 &     0.667\hphantom{*} 	 &     0.667\hphantom{*} \\
\hline\hline
Method features 	 & \#methods 	 & Age 	 & Gender 	 & Age 	 & Gender 	 &   Age 	 &   Gender \\
\hline\hline
HOG 	 & 6 	 & 0.03 	 & -0.01 & 1.15 	 & 1.00 	 &     0.420\hphantom{*} 	 &     0.853\hphantom{*} \\
HOG+COV 	 & 2 	 & 0.10 	 & 0.05 & 1.52 	 & 1.32 	 &     0.333\hphantom{*} 	 &     0.333\hphantom{*} \\
HOG+LBP 	 & 2 	 & 0.06 	 & -0.01 & 1.52 	 & 1.00 	 &     1.000\hphantom{*} 	 &     1.000\hphantom{*} \\
channels 	 & 13 	 & 0.04 	 & 0.01 & 1.31 	 & 1.16 	 &     0.011*            	 &     0.088\hphantom{*} \\
multiple 	 & 5 	 & 0.02 	 & 0.02 & 1.12 	 & 1.14 	 &     0.738\hphantom{*} 	 &     0.690\hphantom{*} \\
pixels 	 & 2 	 & 0.11 	 & 0.03 & 1.86 	 & 1.49 	 &     0.667\hphantom{*} 	 &     0.667\hphantom{*} \\
\hline\hline
Method train set 	 & \#methods 	 & Age 	 & Gender 	 & Age 	 & Gender 	 &   Age 	 &   Gender \\
\hline\hline
INRIA 	 & 21 	 & 0.04 	 & 0.01 & 1.24 	 & 1.10 	 &     0.107\hphantom{*} 	 &     0.435\hphantom{*} \\
INRIA+ 	 & 2 	 & -0.08 	 & -0.02 & 0.90 	 & 1.06 	 &     1.000\hphantom{*} 	 &     1.000\hphantom{*} \\
INRIA/Caltech 	 & 6 	 & 0.10 	 & 0.02 & 1.71 	 & 1.20 	 &     0.011*            	 &     0.152\hphantom{*} \\
\hline\hline
\bf{All methods} 	 & 33 	 & 0.04 	 & 0.01 & 1.31 	 & 1.14 	 &     0.043*            	 &     0.270\hphantom{*} \\
\hline
		\end{tabular}
\end{center}
\caption{Average miss rate differences, ratios and rank-sum tests. Differences and ratios are child-adult, female-male, child/adult, female/male---so positive and higher numbers means worse performance for children and female pedestrians. For rank-sum tests, the p-values are shown. Low p-values mean that the distributions are more likely to be of different median value. p-values under 5\% are followed by a *.}
\label{table:analysis}
\end{table*}

Table~\ref{table:analysis} shows the average differences and ratios between miss rates (child-adult and female-male), as well as the p-values of the Wilcoxon rank-sum test. The statistical test checks whether the distribution of miss rates is the same for both age labels (or both gender labels). Low p-values mean that the distributions are more likely to be of different median values.
The averages and tests are made over the results of multiple methods grouped by their features, classifier, or training data.  The table doesn't show features, classifiers or training sets which appear in less than 2 methods.

The results show that there is bias in terms of age (p=0.043), but the bias in terms of gender is not significant (p=0.270). Looking only at methods using the ``INRIA/Caltech'' dataset combination for training, the age bias is significant (p=0.011) even if the number of samples is only 6. These correspond to the methods RandForest, WordChannels, InformedHaar, SpatialPooling, NAMC, SCCPriors, which all have above-average bias. Bias in gender was close to significant for methods using ``channels'' as the classifier (p=0.088). These methods were highly biased in age as well (p=0.011). 
The most age-biased methods in terms of miss rate difference were deep nets (0.11 difference or 1.86 ratio), but the number of samples was low (2) and so not statistically significant.

\section{Discussion}

The results showed a clear bias of pedestrian detection methods in terms of performance: children had higher miss rates on 82\% of the methods, and on 100\% of the 24 top-performing methods. The difference was 0.04 percentage points on average, but up to 0.07 on methods trained on the INRIA/Caltech datasets. These differences were statistically significant.
Female pedestrians also had higher miss rates than male on 72\% of the methods but the differences were not statistically significant.

\subsection{Why the age bias?}

The fact that the 24 top-performing methods all have higher miss rates on children could mean that high-performing methods are overfitting the distribution of appearance of pedestrians in the dataset, which is itself biased towards adults.
This dataset bias is only natural: for example the demographics of France (home to INRIA) as of 2018 is such that 18\% of the population is 0-14 years old \cite{CIAWorldFactbookFrance}. Children are therefore less likely to be found by random sampling. In addition to that, children are often in school during working hours, which can make their presence in pictures even lower.
The difference in performance could also come from an inherent difficulty of detecting children when compared to adults: children are smaller and pedestrian detectors are known to perform poorly on small bounding boxes \cite{Dollar2012}, they are also more varied walkers since they often run, play, bend down to pick things up from the floor, etc.

\subsection{Biased crash outcomes in AVs?}

One of the most worrying conclusions of these results is that algorithmic bias is likely to gain physical salience in mobile robots such as autonomous vehicles. For example, in our dataset the best performing detector is also one where children's and female pedestrians' miss rates is almost double (1.83) that of adults' and males'. A pedestrian being missed by the perception system will most likely lead to a collision, which means crash outcomes could identically be highly biased.
Additionally, children are actually the kinds of pedestrians we might want to protect even more than others---both emotionally and morally in many ethical theories \cite{Bognar2014}.

\subsection{Choice of algorithm is value-laden}

The results in this paper also show that the choice of pedestrian detection algorithm is value-laden. Choosing a particular algorithm is making an implicit trade-off between efficiency and fairness. In the context of mobile robots such as AVs, this is a trade-off between the safety for some people and for others. This choice should be made carefully. For example, even if the child-adult gap of the best-performing method (PCN) is very high in our dataset, children were still detected more often than the next-best method, and so PCN could still be the best choice for an AV company. The decision would be more complicated if this was not the case.

\subsection{Technical solutions and barriers}

The results in this paper point to a need to obtain more balanced datasets for pedestrian detection and for research in fair detectors. However, there might be several barriers to the dataset fix: there are privacy issues when obtaining street-level datasets with recognizable gender, age, and other features. Additionally, some countries may have regulatory protections against photographing children particularly.
Technical solutions at the algorithm level rather than dataset collection level might have to be preferred for these reasons.

\subsection{Limitations of our study}

The limitations of this study are many: we used a single annotator and a single dataset. Our dataset was urban but not of driving situations, and so how these biases would transfer to the AV context is not straightforward. The choice of dataset was necessary because INRIA is one of the few with considerable children pedestrians and considerable closeness to the pedestrians to allow easy age and gender labelling.

\martim{Importantly as well, gender was operationalized as binary and ascribed by the annotator even though it is an unobservable characteristic on a spectrum. We admit that such a choice of categories could potentially be harmful in the way it perpetuates wrong conceptions of gender. We believe the results are still meaningful as they show potential physical safety biases along personal characteristics.}

Finally, the accident bias that we will see in real AV driving scenarios is most likely going to be different from what we measured in our dataset. This is both because of the use of different perception systems (LIDAR is the most common sensor in AVs) and because AVs will see the same pedestrians over multiple images as the car moves closer, and so might have a higher chance to detect the pedestrian. 
Dataset bias can still be an issue in LIDAR data, and so further investigation on this modality is needed.

{\small
\bibliographystyle{ieee_fullname}
\bibliography{phil}
}

\end{document}